\documentclass[a4paper, twocolumn]{article}
\usepackage{amsmath}
\usepackage[pdftex]{graphicx}
\usepackage{amssymb}
\usepackage{amsthm}
\usepackage[font=small, labelfont=bf]{caption}
\usepackage{subcaption}
\usepackage{float}
\usepackage{wrapfig}
\usepackage[english]{babel}
\usepackage[hyphens]{url}
\usepackage[margin = 0.75in]{geometry}
\usepackage{fancyhdr}
\usepackage[round, authoryear]{natbib}
\usepackage[symbol]{footmisc}
\usepackage{centernot}
\usepackage[affil-it]{authblk} 
\usepackage{lmodern}
\usepackage[table,xcdraw]{xcolor}
\usepackage{abstract}
\usepackage[switch]{lineno}

\title{Stop using root-mean-square error as a precipitation target!}
\author[1,2]{Kieran M.~R.~Hunt}

\affil[1]{Department of Meteorology, University of Reading, Reading, UK}
\affil[2]{National Centre for Atmospheric Sciences, University of Reading, Reading UK}

\begin{document}


\twocolumn[
    \begin{@twocolumnfalse} 
      \maketitle  
       
        \begin{abstract}
            \small 

Root-mean-square error (RMSE) remains the default training loss for data-driven precipitation models, despite precipitation being semi-continuous, zero-inflated, strictly non-negative, and heavy-tailed. This Gaussian-implied objective misspecifies the data-generating process because it tolerates negative predictions, underpenalises rare heavy events, and ignores the mass at zero. We suggest replacing RMSE with the Tweedie deviance, a likelihood-based and differentiable loss from the exponential–dispersion family, with variance function $V(\mu)=\mu^p$. For $1<p<2$ it yields a compound Poisson–Gamma distribution with a point mass at zero and a continuous density over $y>0$, matching observed precipitation characteristics. In this paper, we (i) estimate $p$ from the variance–mean power law and show that precipitation across temporal aggregations is far from Gaussian and that the Tweedie power $p$ increases with accumulation length towards a Gamma limit; and (ii) demonstrate consistent skill gains when training deep data-driven models with Tweedie deviance instead of RMSE. In diffusion-model downscaling over Beijing, Tweedie loss improves wet-pixel MAE and extreme recall ($\sim$0.60 vs 0.50 at the 99th percentile). In ConvLSTM nowcasting over Kolkata, Tweedie loss yields improves wet-pixel MAE and dry-pixel hit rates, with improvements autoregressively compounding over lead time (for MAE, $\sim$2\% at $t{+}1$ growing to $\sim$16\% at $t{+}4$). Because the Tweedie deviance is continuous in $p$, it adapts smoothly across scales, offering a statistically justified, practical replacement for RMSE in precipitation-based learning tasks.
        
        \end{abstract}
        
    \vspace{1cm}
    \end{@twocolumnfalse}
]

\section{Introduction}

\subsection{Precipitation distributions}

Precipitation is a semi-continuous process. At the temporal scales of interest for weather and hydrology, it exhibits a large point mass at zero, strictly non-negative support, and usually a positive skew with heavy upper tails.
These features are inherently non-Gaussian, and a common practical response in the literature is to partition precipitation into an intensity component and an occurrence (or frequency) component. Yet the choice of an appropriate probabilistic model remains contested, with the community having accumulated a remarkably broad set of parametric proposals.

A first class of models treats precipitation as a point process, with storm or cell arrivals modelled by (possibly clustered) Poisson processes. Examples include Poisson or Neyman-Scott formulations for event arrivals and spatiotemporal clustering \citep[e.g.,][]{rodriguez1987,cowpertwait1994,cowpertwait2002,ramesh2012} and studies of the extent to which drop counts deviate from Poisson assumptions \citep[e.g.,][]{kostinski1997}. While these approaches make useful stochastic rainfall generators, they do not in themselves prescribe a suitable distribution for nonzero amounts.

A second class of models treats wet period amounts or time-averaged intensities as approximately Gamma distributed. 
Foundational statistical approaches \citep{thom1958,katz1977,wilks1990} have led to the Gamma distribution being applied across a range of climates and datasets \citep[e.g.,][]{cho2004,segond2006,husak2007,amburn2015,murata2020}. More recently, \citet{martinez2019} gave a physical rationale for the emergence of Gamma-like distributions under temporal averaging of rainfall events.

Many alternatives have been explored beyond the Poisson and Gamma distributions. Mixtures of exponentials have been found to fit nonzero daily amounts better than a single Gamma in certain regions \citep[e.g.,][]{wilks1998,wilks1999,woolhiser1982,rho2019}. Lognormal models have also been proposed, particularly for satellite-derived rain rates \citep[e.g.,][]{cho2004,gupta1998}. Other families such as Weibull \citep[e.g.,][]{wilks1989,olivera2019} and kappa distributions \citep[e.g.,][]{strong2025} have also been proposed, and non-parametric or flexible parametric approaches have been explored for cases when a single family seems too restrictive.
For extremes, generalised extreme‑value (GEV) and generalised Pareto distributions are widely used to represent annual (or other timescale) maxima and threshold exceedances \citep[e.g.,][]{papalexiou2013,aghakouchak2010,phoophiwfa2024}.

Because of the difficulty in asserting a single parametric form for precipitation, many evaluation studies deliberately avoid strong distributional assumptions, instead relying on non-parametric or distribution-agnostic scores. For verification of spatial structure, e.g., for gridded rainfall, neighbourhood or scale-aware methods such as the fractions skill score are common \citep[e.g.,][]{ebert2004,roberts2008,mittermaier2021}.
For event occurrences, categorical scores (e.g., the Brier score) are popular \citep[e.g.,][]{brier1950,hogan2010}. For continuous probabilistic assessments, proper scores such as the continuous ranked probability score (CRPS) and feature-based metrics such as structure-amplitude-location (SAL) are widely used, especially in validation of operational forecasts \citep[e.g.][]{hersbach2000,wernli2008}.

\subsection{The problem with mean-square error}

Such metrics are very useful for model intercomparison and process-level diagnostics; however, they do not address the central issue in training data-driven models -- that one must choose a differentiable loss function that somehow encodes the assumed data-generating mechanism.
In practice, most deterministic regression models for precipitation are trained with root-mean-square error or mean-square error (RMSE/MSE; hereafter we will simply refer to both as MSE). This implicitly assumes Gaussian residuals with symmetric errors, which is violated by precipitation. As a result, models trained on MSE under-penalise failures on rare heavy events, allow predictions of negative rainfall (which must be corrected post hoc), and mishandle the mass at zero.

Despite these obvious shortcomings, MSE remains in near-ubiquitous use for data-driven precipitation models. For example, in nowcasting \citep[e.g.,][]{lebedev2019,xiang2020,cambier2023,han2023,kim2024,ayzel2025,cao2025}. In statistical and data-driven downscaling, MSE is similarly standard as the pixel-wise training objective \citep[e.g.,][]{vandal2017,kumar2021,wang2021b,reddy2023,chiang2024,liu2025}.
Other precipitation-related learning tasks (e.g., merging datasets, data-driven discovery, and AI-based global weather forecast models) also default to MSE \citep[e.g.,][]{moraux2021,hunt2024,lam2023,kochkov2024}. 
A few studies have begun to explore alternatives, e.g. frequency-domain or correlation-aware losses for nowcasting \citep{yan2024}, cross-entropy when modelling occurrences \citep[e.g.,][]{shi2015,agrawal2019}, or adversarial/transport‑style objectives in downscaling \citep[e.g.,][]{kumar2023}. However, these remain the exception rather than the rule.

\subsection{Tweedie distributions}
\label{sec:tweedie-literature}

Against this backdrop, the Tweedie family offers an attractive compromise. As an exponential‑dispersion family with variance function $V(\mu)=\mu^p$, Tweedie models provide a continuum connecting Gaussian ($p\to0$), Poisson ($p\to1$), and Gamma ($p\to2$) limits, and for $1<p<2$ represent a compound Poisson–Gamma law that accommodates zero inflation together with continuous, positively supported amounts.
This makes the Tweedie particularly appealing for precipitation, where we want to simultaneously model a frequent no-rain state alongside a skewed distribution of strictly positive intensities.
Earlier methodological work has made likelihood‑based fitting practical \citep[e.g.,][]{dunn2004}, and applications to monthly and seasonal rainfall in different climates have demonstrated competitive performance and interpretability \citep[e.g.,][]{hasan2010,hasan2011,hasan2012,svensson2017,suhaila2023,dzupire2018}. Nevertheless, Tweedie models remain underused relative to their relevance.

In this paper, we will argue that the training objective must be statistically consonant with the target variable. Because precipitation is non-negative, highly skewed, and frequently zero, a Gaussian-implied MSE is a poor default. In contrast, the Tweedie deviance (derived in Section \ref{sec:tweedie}) provides a coherent likelihood-based loss that accounts for these characteristics while remaining tractable for gradient-based training. Crucially, the Tweedie index parameter $p$ can be estimated empirically (e.g., from the variance-mean power law) and adapted to different temporal aggregations, yielding a family of training targets that correctly interpolate between Poisson-like and Gamma-like behaviours rather than incorrectly imposing an a priori Gaussian assumption.
The paper is laid out as follows. 
In Sec.~\ref{sec:tweedie}, we explain in more detail the features of the Tweedie that make it desirable. We then demonstrate that Gaussianity is not a reasonable baseline for precipitation distributions across scales and show how to estimate the Tweedie power parameter from data (Sec.~\ref{sec:goodness-of-fit}), before quantifying the impact of replacing MSE with Tweedie deviance in twp common learning tasks for precipitation: downscaling (Sec.~\ref{sec:downscaling}) and nowcasting (Sec.~\ref{sec:nowcasting}).

\section{Important features of the Tweedie distribution}
\label{sec:tweedie}

\begin{figure*}[htbp]
\centering
\includegraphics[width=.75\linewidth]{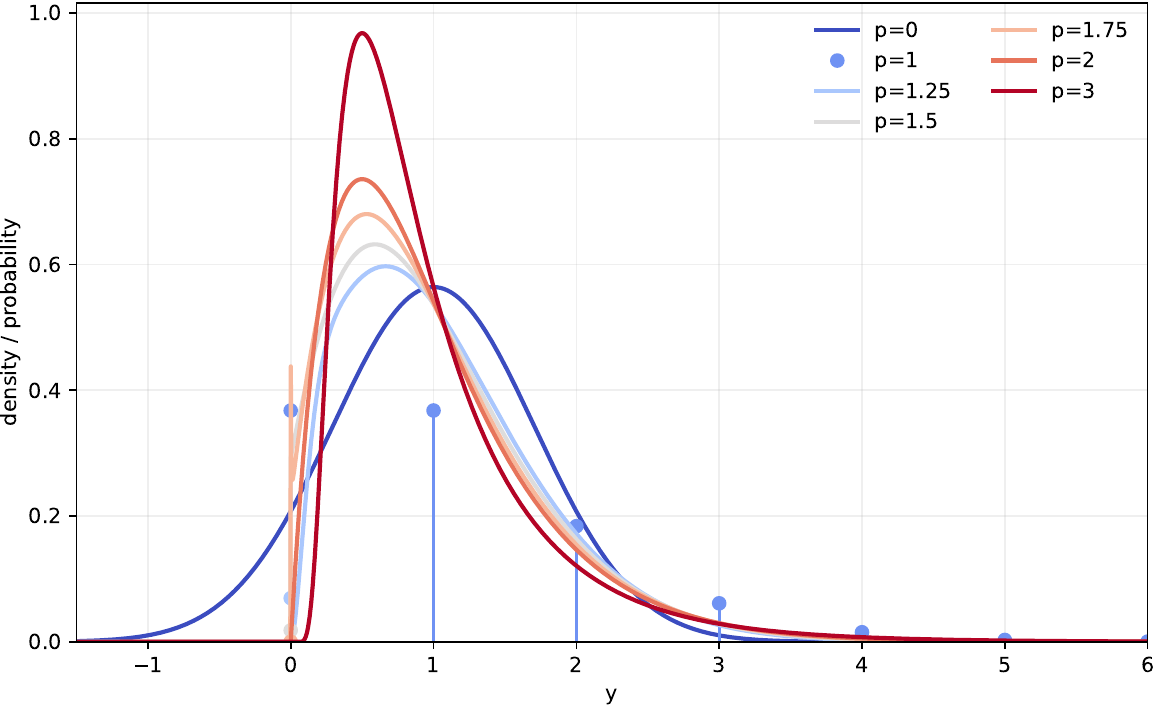}
\caption{Tweedie family PDFs for fixed mean and dispersion. Curves show $f_Y(y)$ for $\mu=1$ and $\phi=0.5$ across powers $p\in\{0,1,1.25,1.5,1.75,2,3,4\}$. Special cases: $p=0$ is $\mathcal N(\mu,\phi)$ (solid line), $p=1$ is Poisson$(\mu)$ (probability mass function shown as stems), $p=2$ is Gamma, $p=3$ is inverse-Gaussian. For $1<p<2$ the distribution is compound Poisson–Gamma with a point mass at zero (stem height $\exp\{-\mu^{\,2-p}/[\phi(2-p)]\}$) and a continuous density on $y>0$. For $p>1$ support is $y>0$; the variance scales as $\mathrm{Var}(Y)=\phi\,\mu^{p}$. Continuous lines are densities; stems indicate discrete probability mass.
}
\label{fig:tweedie-family}
\end{figure*}

An introduction to the Tweedie distribution and its advantages follows.
Throughout, we will use $y$ to mean the observation, $\theta$ to mean the canonical parameter of the distribution, $\kappa$ to mean the cumulant function, and $\phi>0$ to mean the dispersion.
For any power $p$ (with $p\neq1,2$) the Tweedie belongs to the exponential dispersion family:
\begin{equation}
\label{eq:tweedie-pdf}
    f_{Y}(y;\theta,\phi)
	= a(y,\phi,p)\;
	  \exp\left[\frac{y\theta-\kappa(\theta)}{\phi}\right],
\end{equation}
whose variance function is
\begin{equation}
    \operatorname{Var}(Y)=\phi\,V(\mu),\qquad V(\mu)=\mu^{p}.
\end{equation}
The mean $\mu$ and canonical parameter $\theta$ are then related through:
\begin{equation}
\label{eq:theta-kappa}
    \theta = \frac{\mu^{1-p}}{\,1-p\,},\qquad 
\kappa(\theta) = \frac{\mu^{2-p}}{\,2-p\,},
\end{equation}
and the normalising factor, $a(\cdot)$, is whatever constant in $y$ is required to make the density function integrate to one.

What we are really interested in, however, is the deviance. 
For a single observation, $y$, the unit deviance is the scaled log-likelihood ration between the saturated (i.e., perfectly fit) model and model with mean $\mu$:
\begin{equation}
    d(y,\mu)=2\left[ \ell(y;\theta_y,\phi)-\ell(y;\theta_\mu,\phi)\right],
\end{equation}
where 
\begin{equation}
    \ell(y;\theta,\phi) = \dfrac{y\theta-\kappa(\theta)}{\phi}+ \log a(y,\phi,p)\, ,
\end{equation} 
and $\theta_y$ is the canonical parameter corresponding to mean $y$ (similarly for $\mu$).
Because the normalising term $\log a(y,\phi,p)$ does not depend on $\theta$, it cancels in the difference, and the unit deviance (i.e.~without the $1/\phi$ factor) is
\begin{equation}
\label{eq:log-lik}
d(y,\mu)=2\left[y\left(\theta_y-\theta_\mu\right)-\left(\kappa(\theta_y)-\kappa(\theta_\mu)\right)\right]\, .
\end{equation}
The difference terms can be computed through integration, i.e.:
\begin{equation}
    \theta_y-\theta_\mu=\int_{\mu}^{y}\frac{d\theta}{d t}dt=\int_{\mu}^{y}\frac{dt}{V(t)}
\end{equation}
and
\begin{equation}
    \kappa(\theta_y)-\kappa(\theta_\mu)=\int_{\theta_\mu}^{\theta_y}\kappa'(\theta)d\theta=\int_{\mu}^{y}\frac{t}{V(t)}dt\, ,
\end{equation}
where we have used the identity $\mu = \mathbb E(y)=\kappa'(\theta)$ which holds for all members of the exponential distribution family.
Substituting these values into Eq.~\ref{eq:log-lik} then gives a single integral form for the unit deviance of any exponential distribution function:
\begin{equation}
\label{eq:edf-deviance}
    d(y,\mu)=2\int_{\mu}^{y}\frac{y-t}{V(t)}dt\, .
\end{equation}
Thus, for the Tweedie, we have:
\begin{equation}
\label{eq:tweedie-deviance}
\begin{aligned}
    d_p(y,\mu)&=2\int_{\mu}^{y}(y-t)\,t^{-p}\,dt \\
&=2\left[
y\int_{\mu}^{y} t^{-p}dt-\int_{\mu}^{y} t^{1-p}dt
\right]\\
&= 2\left[\frac{y^{1-p}-\mu^{1-p}}{1-p}-\frac{y^{2-p}-\mu^{2-p}}{2-p}\right]\\
&=2\left[\frac{y^{\,2-p}}{(1-p)(2-p)}-\frac{y\,\mu^{\,1-p}}{1-p}+\frac{\mu^{\,2-p}}{2-p}\right],
\end{aligned}
\end{equation}
for $p\neq1,2$. The total deviance for a sample, $D$, is then obtained in the usual way, $D=\frac{1}{\phi}\Sigma_i d_p(y_i,\mu_i)$.

\subsection{Recovering the Gaussian ($p\to0$)}
Let $p=\varepsilon$ with $\varepsilon\to0$. We then expand Eq.~\ref{eq:theta-kappa} as Taylor series:
\begin{equation}
\begin{aligned}
    \theta &= \frac{\mu^{1-\varepsilon}}{1-\varepsilon}
	= \mu\bigl[1-\tfrac{\varepsilon}{2}(\ln\mu)\bigr]+O(\varepsilon^{2}),\\
\kappa(\theta)&=\frac{\mu^{2-\varepsilon}}{2-\varepsilon}
	=\tfrac12\mu^{2}+O(\varepsilon).
\end{aligned}
\end{equation}
Keeping only leading terms and substituting back into Eq.~\ref{eq:tweedie-pdf}, we obtain the expected probability density function:
\begin{equation}
    f_{Y}(y;\mu,\phi) \;\longrightarrow\;
\underbrace{(2\pi\phi)^{-1/2}}_{a(y,\phi,0)}
\exp\left[-\tfrac{(y-\mu)^{2}}{2\phi}\right],
\end{equation}
where $a$ is determined through integration.

The Gaussian deviance, with $p\to0$ and $V(\mu)=1$, is then given by Eq.~\ref{eq:edf-deviance}:
\begin{equation}
    d_{0}(y,\mu)=2\int_{\mu}^{y}(y-t)\,dt=(y-\mu)^2 \, ,
\end{equation}
which is, of course, the mean squared error.

\subsection{Recovering the Poisson ($p\to1$)}
Let $p=1+\varepsilon$, again with $\varepsilon\to0$. Then
\begin{equation}
\begin{aligned}
    \theta &= \frac{\mu^{-\varepsilon}}{-\varepsilon}
	= \ln\mu + O(\varepsilon)\\
\kappa(\theta) &= \frac{\mu^{1-\varepsilon}}{1-\varepsilon}
	= \mu + O(\varepsilon).
    \end{aligned}
\end{equation}
Keeping only leading terms and substituting back into Eq.~\ref{eq:tweedie-pdf}, we obtain the probability density function:
\begin{equation}
    f_{Y}(y;\mu) = \underbrace{\frac{1}{y!}}_{a(y,\phi,1)}\mu^{\,y}\,e^{-\mu} \, .
\end{equation}

To obtain the deviance, we again use Eq.~\ref{eq:edf-deviance}:
\begin{equation}
    d_{1}(y,\mu)=2\int_{\mu}^{y}\frac{y-t}{t}\,dt
  =2\left[y\ln(y/\mu)-(y-\mu)\right] \, ,
\end{equation}
which is, of course, not the mean squared error.

\subsection{Recovering the Gamma ($p\to2$)}
Let $p=2-\varepsilon$ with $\varepsilon\to0$. From Eq.\ref{eq:tweedie-pdf}, we have:
\begin{equation}
\begin{aligned}
    \theta &= \frac{\mu^{-(1-\varepsilon)}}{-(1-\varepsilon)}
	= -\frac{1}{\mu}+O(\varepsilon)\\
\kappa(\theta)&=\frac{\mu^{\varepsilon}}{\varepsilon}
	= \ln\mu + O(\varepsilon).
\end{aligned}
\end{equation}
The normalising constant is rather more complicated than for $p=0$ or $p=1$:
\begin{equation}
    a(y,\phi,2)=\frac{y^{\frac1\phi-1}}{\Gamma(\frac1\phi)\,\phi^{\frac1\phi}}\, ,
\end{equation}
giving 
\begin{equation}
    f_{Y}(y;\mu,\phi)
	=\frac{1}{\Gamma(\tfrac1\phi)}
	\Bigl(\frac{y}{\phi\mu}\Bigr)^{\frac1\phi-1}
	\frac{e^{-y/(\phi\mu)}}{\mu} \,.
\end{equation}
The gamma distribution is more commonly written in terms of a shape parameter, $k=\frac1\phi$, and a scale parameter, $\beta=\tfrac{\mu}{k}=\phi\mu$. Thus we return the commonly known density function:
\begin{equation}
    \operatorname{Gamma}(y;\;k,\beta)=
\frac{y^{k-1}e^{-y/\beta}}{\Gamma(k)\,\beta^{k}}\, .
\end{equation}

Returning to Eq.~\ref{eq:edf-deviance}, the deviance is then given by:
\begin{equation}
    d_{2}(y,\mu)=2\int_{\mu}^{y}\frac{y-t}{t^{2}}\,dt
  =2\left[\frac{y-\mu}{\mu}-\ln\!\left(\frac{y}{\mu}\right)\right]
\end{equation}
which is, of course, not the mean squared error.

\section{Data and models}
\subsection{Sources of precipitation data}
\subsubsection{University of Reading observatory}

We use rain–gauge observations from the University of Reading Atmospheric Observatory on the Whiteknights campus (United Kingdom). The site records precipitation with a tipping-bucket gauge connected to a METFiDAS-3 data logger, and provides WMO-standard processed output at 5-minute resolution from September 2014 onwards (details and additional instrumentation are described at \url{https://www.met.reading.ac.uk/observatorymain/precipitation.html}). 

For this study, we work with the 5-minute series in units of mm per interval. Basic quality control is applied to ensure physical values: negative amounts are set to zero and non-finite entries are treated as missing. No wind-related undercatch correction or bias adjustment is applied. When accumulating to coarser temporal scales (30-min, hourly, daily, weekly, monthly), totals are formed by summation of the 5-min amounts within each interval. Intervals with missing sub-samples are excluded. 

These gauge data provide a point-scale benchmark with high zero frequency at short accumulation times, which is we will use for assessing the suitability of Tweedie-family losses and for contrasting behaviour across temporal aggregation (Section~\ref{sec:goodness-of-fit}).
Data can be freely downloaded from \url{https://metdata.reading.ac.uk/cgi-bin/MODE3.cgi}.

\subsubsection{GPM-IMERG}
GPM-IMERG \citep[Global Precipitation Mission—Integrated Multi-satellitE Retrievals for GPM;][]{huffman2015} has global coverage at a half-hourly, 0.1° resolution, starting June 2000 and continuing to the present day. Over the tropics, GPM-IMERG primarily ingests retrievals from (for 2000–-2014) the defunct Tropical Rainfall Measuring Mission \citep{kummerow1998,kummerow2000} 13.8~GHz precipitation radar and microwave imager \citep{kozu2001} and (for 2014--) the Global Precipitation Measurement \citep[GPM;][]{hou2014} Ka/Ku-band dual-frequency precipitation radar. When an overpass is not available, precipitation is estimated by calibrating infrared measurements from geostationary satellites. While GPM IMERG performs well when compared against gauge-based products, performance falls at higher elevations or when quantifying extreme precipitation events \citep{prakash2018}. Data were downloaded from \url{https://disc.gsfc.nasa.gov/datasets/GPM_3IMERGHH_06/summary?keywords=%22IMERG%20final%22}.

\subsubsection{ERA5}
The ECMWF ERA5 reanalysis \citep{hersbach2020} provides global coverage at hourly resolution from 1940 to present on a $0.25^\circ$ grid, with 37 pressure levels from 1000 to 1~hPa, plus single-level (surface, near-surface, or column-integrated) variables. It assimilates a wide range of conventional and satellite observations via 4D-Var. Many hydrometeorological quantities (including precipitation) are model diagnostics produced by the forecast model within the reanalysis rather than being directly assimilated.

For this study we use the single-level total precipitation field. Further preprocessing details (domain, splits, coarsening, thresholds) are given in App.~\ref{app:downscale}. Data were downloaded using the dedicated API \url{https://cds.climate.copernicus.eu/cdsapp#!/dataset/reanalysis-era5-pressure-levels}.

\subsection{Outline of model architectures}

\subsubsection{Downscaling}
\label{sec:downscaling-method}
We train a conditional diffusion model to map coarse (1$^\circ$) precipitation to the original 0.25$^\circ$ ERA5 precipitation. The score network is a U--Net with group normalisation and sinusoidal time embeddings. Model input is the high-resolution noise-added field concatenated with (i) the upsampled coarse precipitation and (ii) a static orography channel. We follow a linear noise schedule with $T{=}1000$ steps and sample with a deterministic denoising diffusion implicit model scheme \citep{song2020}. Targets are non-negative, so we pass the network output through a softplus to enforce $\mu\ge0$. 
We then compare loss functions on exactly the same architecture: (a) RMSE on the target $x_0$ (i.e., the clean 0.25° precipitation field) ; and (b) Tweedie deviance (Sec.~\ref{sec:tweedie}) with power $p$ estimated from the training set via a variance–mean power law (Sec.~\ref{sec:goodness-of-fit}) on the same target. More granular details of the setup are given in App.~\ref{app:downscale}.

\subsubsection{Nowcasting}
\label{sec:nowcasting-method}
For 30-min IMERG nowcasting over the Kolkata domain we use a residual ConvLSTM \citep{shi2017}. This comprises two stacked convolutional LSTM blocks with a skip connection, batch normalisation, and a shallow (i.e., single layer) convolutional head. As in the downscaling model, we use softplus to ensure non-negativity. Inputs are the previous $L{=}4$ frames; the model predicts the next frame, and longer horizons are obtained by running the model autoregressively. As before, we train the same architecture with different losses: (a) RMSE and (b) Tweedie deviance. More granular details of the setup are given in App.~\ref{app:nowcast}.

\section{Case study applications}
We will now present three case studies. The first will demonstrate that the Gaussian is not a reasonable \emph{a priori} assumption for the distribution of precipitation.
The second and third will demonstrate that nontrivial gains can be made in the skill of models used for common tasks (downscaling and nowcasting respectively) when the loss function is changed from RMSE to an appropriately constrained Tweedie deviance.

\subsection{Goodness-of-fit to observations}
\label{sec:goodness-of-fit}

\begin{figure*}[htbp]
\centering
\includegraphics[width=\linewidth]{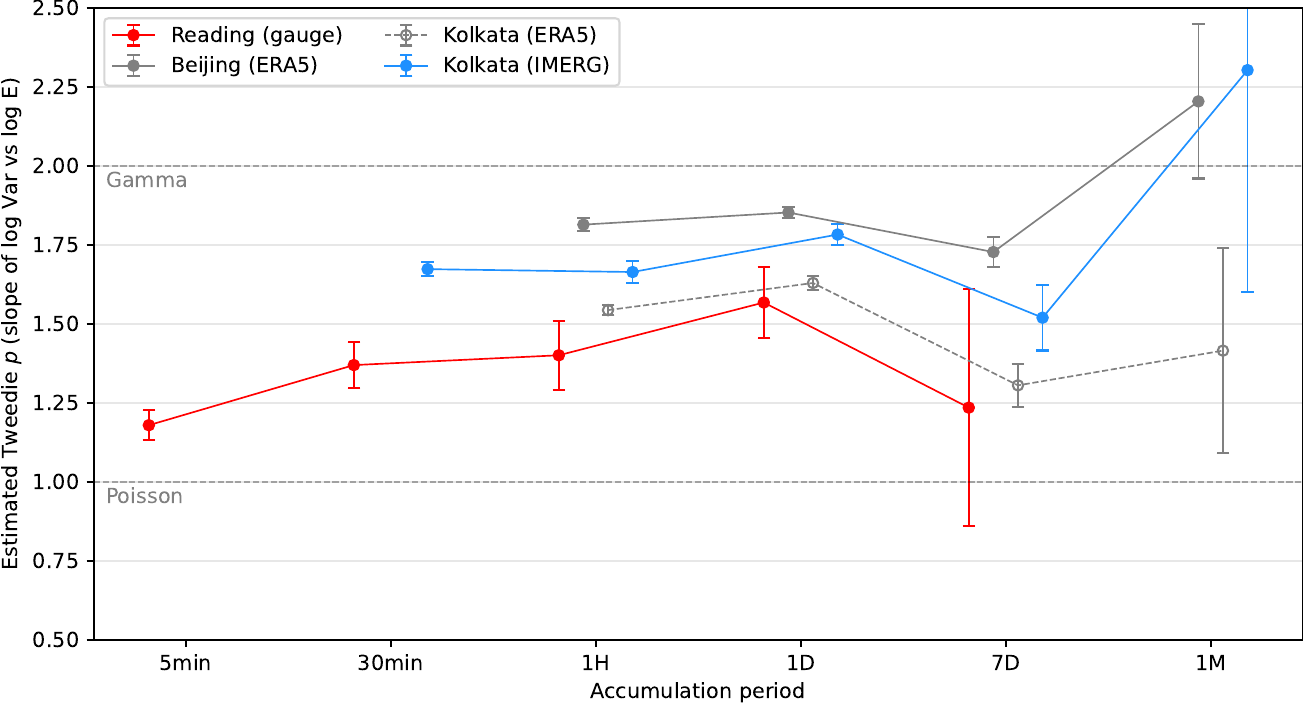}
\caption{Estimated Tweedie power parameter $p$ as a function of accumulation period for four datasets: Reading gauge (red), ERA5 at Beijing (grey), ERA5 at Kolkata (dashed grey) and IMERG at Kolkata (blue). For each dataset and accumulation period, totals are formed by summing raw values over the period and $p$ is estimated from the power–law relation $\log\operatorname{Var}(Y)=c+p\,\log\operatorname{E}[Y]$ using ordinary least squares on non-overlapping blocks. Block lengths are: 10 days (5-min gauge), 15 days (30-min series), 30 days (hourly and daily), 16\,weeks (weekly), and 12 months (monthly). Error bars show 95\% Wald confidence intervals from the ordinary least squares slope standard error.}
\label{fig:tweedie-p-by-aggregation}
\end{figure*}

Figure~\ref{fig:tweedie-p-by-aggregation} summarises how the Tweedie power parameter, $p$, varies with accumulation period across our three independent datasets. A clear trend emerges, in that at short accumulations the process is close to the Poisson limit ($p\to1$), while longer accumulations progressively drive $p$ upwards towards the Gamma limit ($p\to2$). In particular, the 5-min gauge data from Reading is very close to a Poisson distribution ($p=1.18$), whereas the monthly totals in both ERA5 and IMERG have confidence intervals overlapping $p=2$.

At the same time, there is a notable spread between datasets at the same accumulation period. For example, for daily totals, the estimated powers span roughly $p\in[1.57,1.85]$. This inter-dataset variability is smaller than the systematic increase with accumulation period, but it is not negligible and cautions us against assuming a single $p$ for a given timescale.

Including ERA5 for Kolkata shows the additional importance of location. Even with the same preprocessing, the daily estimates differ considerably between sites (Kolkata $p=1.629$, Beijing $p=1.852$). This partly reflects differences in storm organisation and frequency (e.g., convective storms embedded in the summer monsoon versus mixed regimes in the mid-latitudes), while the difference between IMERG and ERA5 at Kolkata indicates that dataset characteristics also contribute.

There are two practical conclusions from this. Firstly, none of the precipitation series shown here are remotely Gaussian at any accumulation: a Gaussian-consistent loss would require $p\to0$, whereas all estimates sit in the Poisson-Gamma regime ($1<p<2$). Secondly, $p$ is not knowable \emph{a priori}. Instead, it must first be estimated from the training data at the target temporal scale and carried through model training and evaluation so that the deviance matches the problem.

Finally, we note that weekly accumulations consistently have lower $p$ than daily accumulations. This holds for all datasets, but it is unclear why. Further sensitivity tests are required.

\subsection{Use in training a downscaling model}
\label{sec:downscaling}

\begin{figure*}[htbp]
\centering
\includegraphics[width=\linewidth]{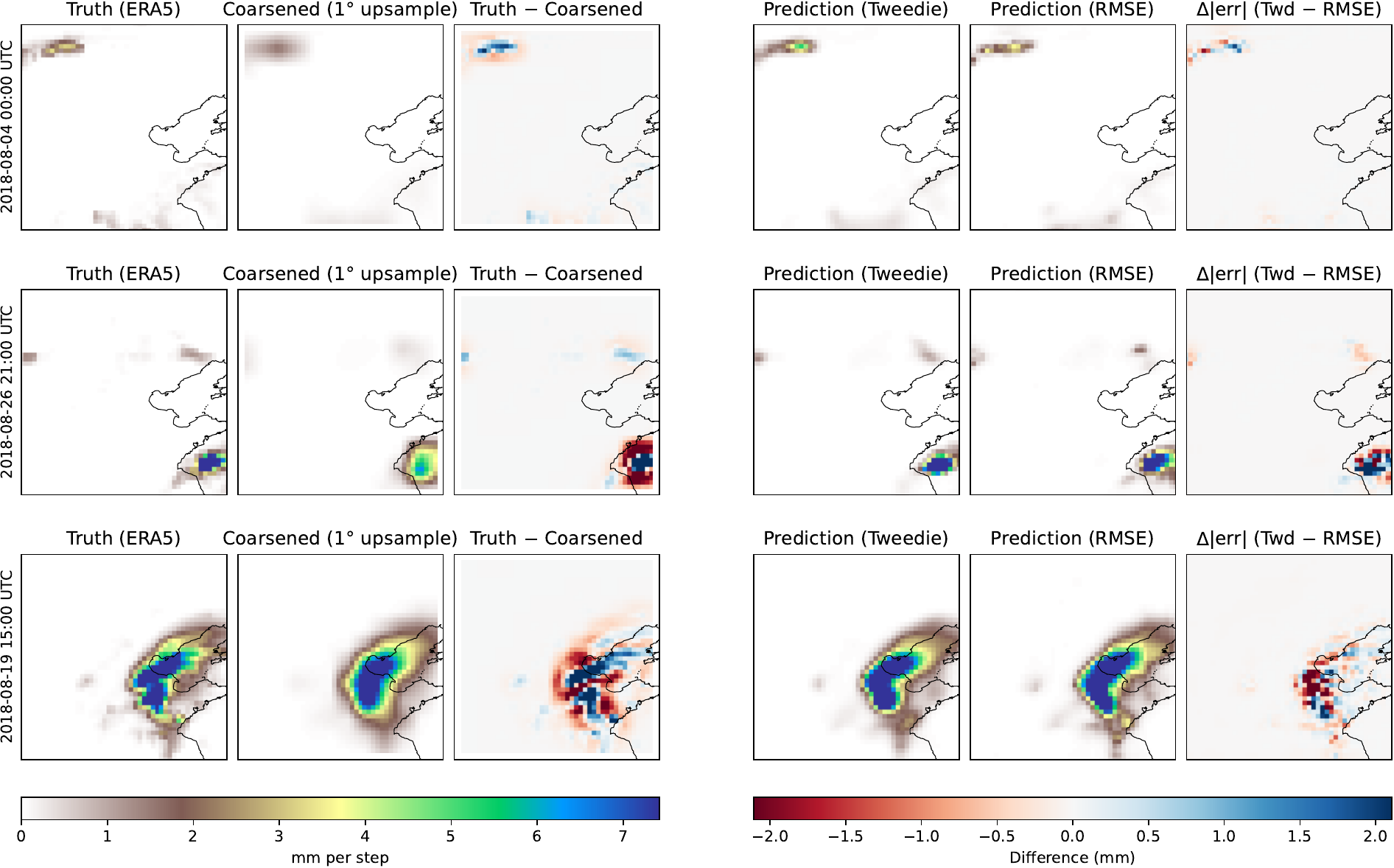}
\caption{Performance of Tweedie vs RMSE diffusion downscaling for three randomly selected test cases. Column 1 shows the actual precipitation in ERA5; column 2 shows the ERA5 precipitation coarsened to 1°, i.e., the model input; and column 3 shows their difference. Column 4 and 5 show the predictions from the Tweedie and RMSE models respectively; and and column 6 shows the difference in the mean absolute error between the models such that red (i.e., negative) values indicating a smaller error in the Tweedie model.}
\label{fig:downscale-cases}
\end{figure*}

\begin{figure*}[htbp]
\centering
\includegraphics[width=\linewidth]{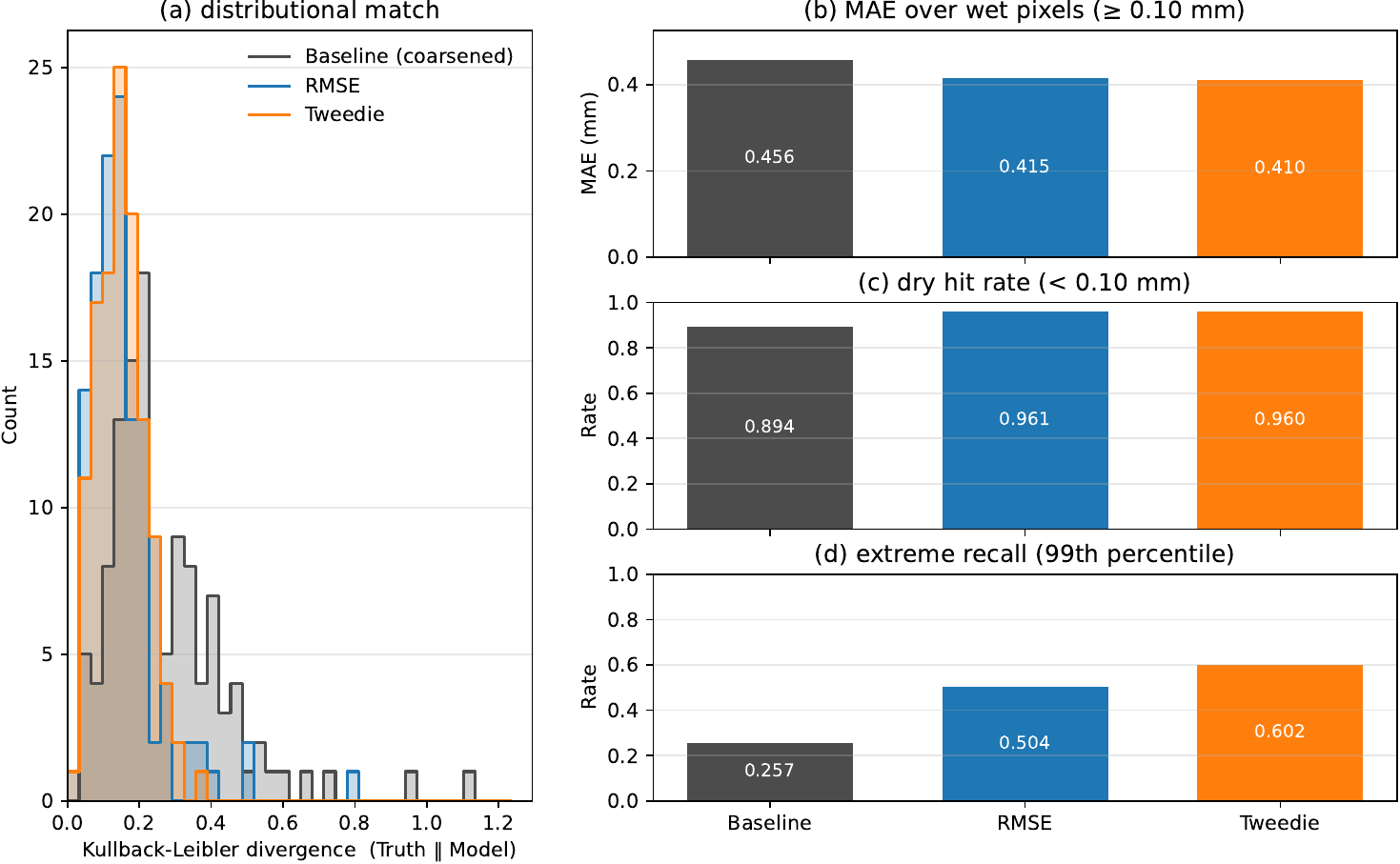}
\caption{Comparison of Tweedie vs RMSE diffusion downscaling over the August 2018 Beijing test period.
(a) Distribution of per-timestep Kullback–Leibler divergence $D_{\mathrm{KL}}(\text{Truth}\parallel\text{Model})$ computed from histogrammed pixels (with ERA5 as truth).
(b) mean absolute error over wet pixels (truth $\ge$ 0.1 mm hr$^{-1}$);
(c) dry hit rate $= \Pr(\text{pred} < 0.1\ \text{mm hr}^{-1} \mid \text{truth} < 0.1\ \text{mm hr}^{-1})$;
(d) extreme recall at the 99th-percentile (computed from truth over the period).
Blue bars/curves show the RMSE-trained model; orange shows the Tweedie-trained model; grey shows the scores achieved by simply using the 1° coarsened data (i.e., no model) as a baseline. All metrics average over valid grid cells/times.}
\label{fig:downscale-metrics}
\end{figure*}

We now consider the relative performance of the downscaling model (Sec.~\ref{sec:downscaling-method}) trained to minimise RMSE and Tweedie deviance respectively.
Randomly-selected precipitating case studies from August in our testing period (Fig.~\ref{fig:downscale-cases}) reveal mixed results. The first case (top row) has light precipitation in the northwest of the domain, which the Tweedie model overestimates and the RMSE model underestimates. In the second case, a small but more intense storm in the southeast of the domain, both models correctly predict the high intensity and storm morphology, but the RMSE model has, in general, a lower mean absolute error. In the third case, a very intense and large storm, both models again predict the intensity and storm morphology correctly, but the Tweedie model has a lower overall absolute error.

Extending the analysis to the whole month of August 2018 (Fig.~\ref{fig:downscale-metrics}), we see that the two models perform similarly. We first compute, for each output, the Kullback-Leibler divergence between the output and the `true' ERA5. The two distributions are virtually indistinguishable, and are not significantly different under a two-sample Kolmogorov-Smirnov test.
This also holds for the MAE on wet pixels (RMSE: 0.415; Tweedie: 0.410) and the dry hit rate (RMSE: 0.961; Tweedie: 0.960). Again, the differences are not statistically significant when bootstrap tested.
However, we do find a significant difference in 99th percentile recall, which is 0.504 for the RMSE model and 0.602 for the Tweedie model, a significant improvement.

So, although the two downscaling models have very similar performance across a range of metrics, we see a 19\% improvement in predicting extreme rainfall with the Tweedie model over the RMSE model, due to better representation of the right tail in the loss function of the former.


\subsection{Use in training a nowcasting model}
\label{sec:nowcasting}

\begin{figure*}[htbp]
\centering
\includegraphics[width=\linewidth]{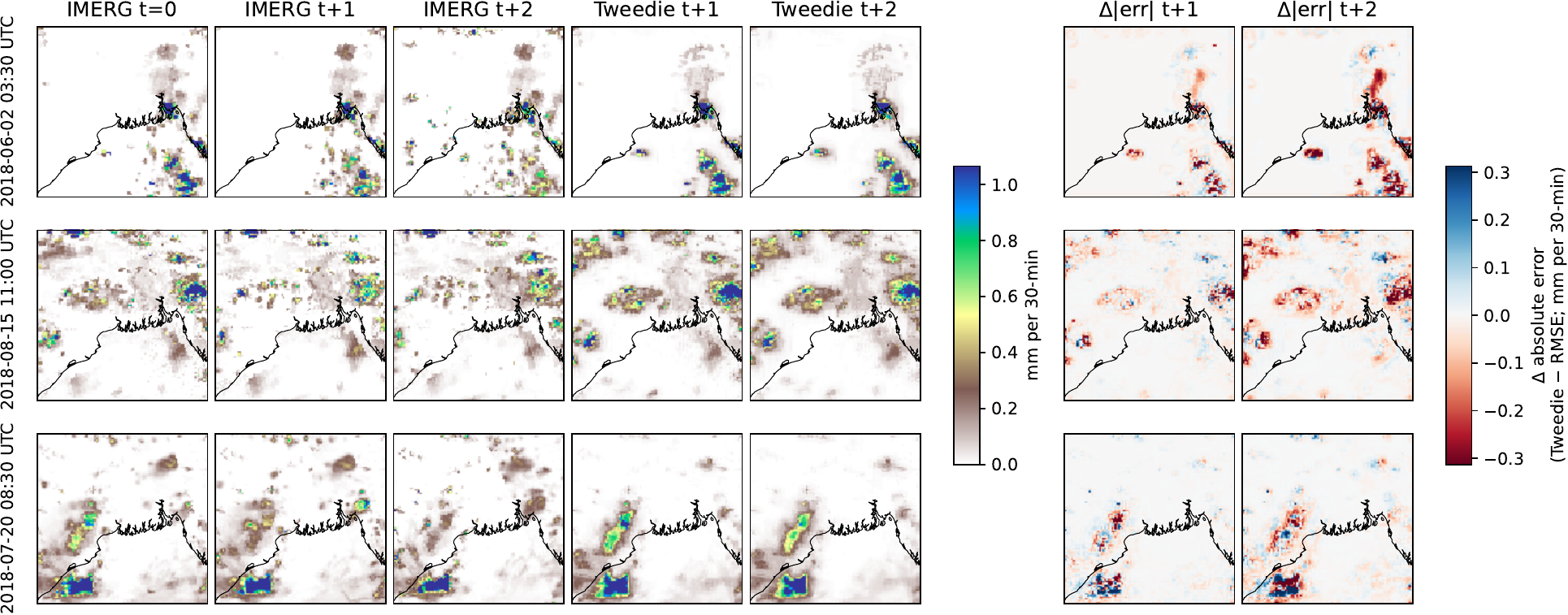}
\caption{Performance of Tweedie vs RMSE ConvLSTM Kolkata nowcasts for three randomly selected test initialisations. Column 1 shows the last observed IMERG frame (t=0) that the models see; columns 2–3 show the subsequent half-hourly observations (t+1, t+2). Columns 4–5 show Tweedie nowcasts at t+1 and t+2 (with t+2 obtained autoregressively by feeding the t+1 prediction back as input).
Columns 6-7 show the difference in mean absolute error such that red (i.e., negative) values indicating a smaller error in the Tweedie model.}
\label{fig:nowcasting-cases}
\end{figure*}

\begin{figure*}[htbp]
\centering
\includegraphics[width=\linewidth]{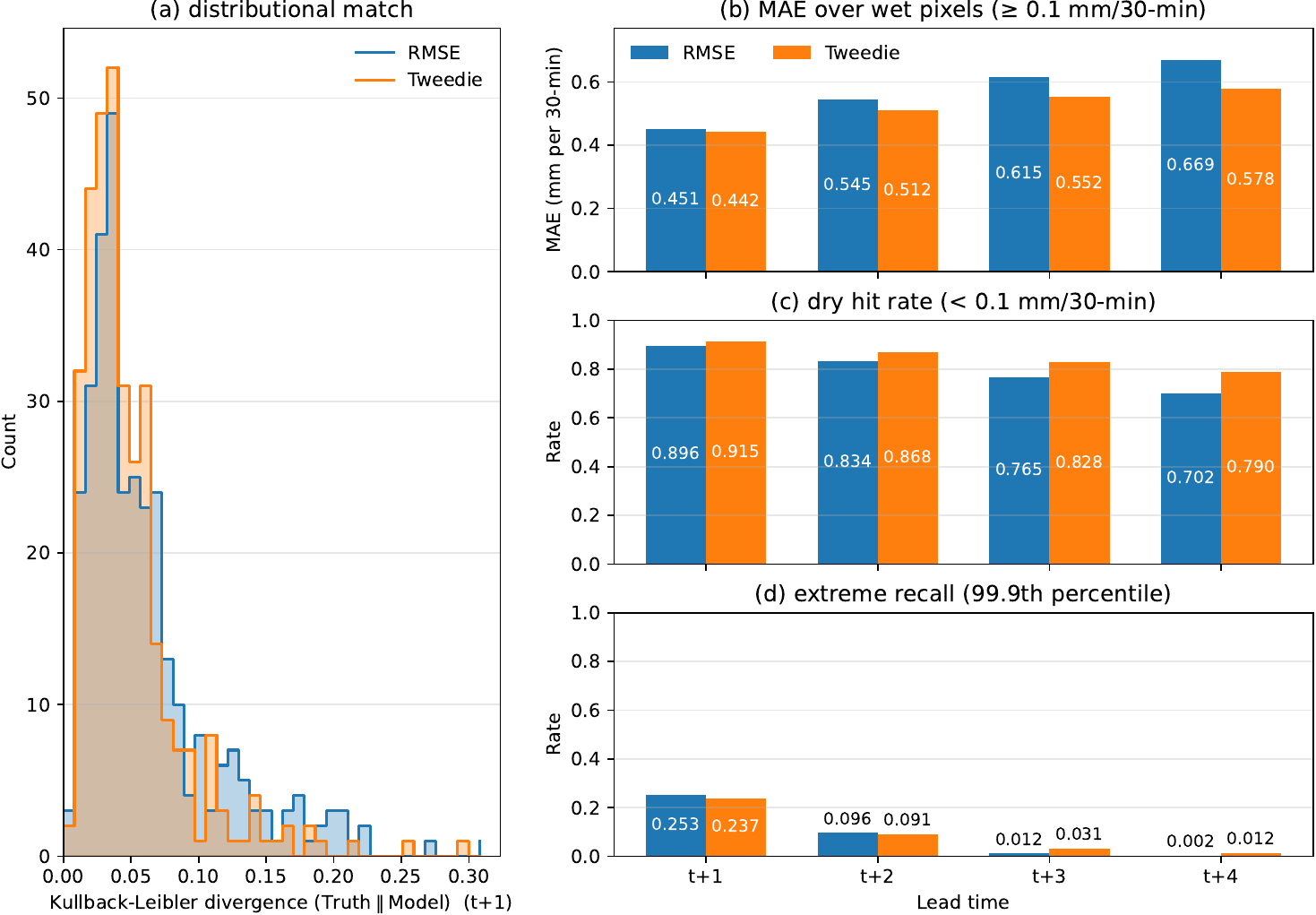}
\caption{Comparison of Tweedie vs RMSE ConvLSTM nowcasts over the August 2018 Kolkata test period.
(a) Distribution of t+1 Kullback–Leibler divergence $D_{\mathrm{KL}}(\text{Truth}\parallel\text{Model})$ with IMERG as truth.
(b–d) Mean scores vs lead time (t+1 to t+4) from autoregressive rollouts (each step feeds the previous prediction back as input):
(b) mean absolute error over wet pixels (truth $\ge$ 0.1 mm per 30 min);
(c) dry hit rate $= \Pr(\text{pred} < 0.1\ \text{mm} \mid \text{truth} < 0.1\ \text{mm})$;
(d) extreme recall at the 99.9th-percentile threshold (computed from truth over the period).
Blue bars/curves show the RMSE-trained model; orange shows the Tweedie-trained model. All metrics average over valid grid cells/times.
}
\label{fig:kl_mae_combo}
\end{figure*}

Finally, we consider the respective performances of the Tweedie- and RMSE-trained nowcasting models, described in Sec.~\ref{sec:nowcasting-method}. Case study maps for a four-timestep rollout (Fig.~\ref{fig:nowcasting-cases}) show that the Tweedie model is able to capture the structural features of the rainfall field, as well as broad characteristics of its evolution, including movement, growth and decay, across a range of scales.
We are interested in the relative performance of the two models, shown in the rightmost two columns of Fig.~\ref{fig:nowcasting-cases}. 
For both timesteps, and across all three cases, we see that the Tweedie model has lower error than the RMSE model at almost all locations. This includes both heavy rain and light rain events -- although we note that for the heaviest rainfall case (bottom row), RMSE and Tweedie have a roughly balanced performance.

We evaluate the model performances over the same test month (August 2018; Fig.~\ref{fig:kl_mae_combo}), chosen to cover some of the summer monsoon.
We first compute, for each $t+1$ output initialised from within the test period, the KL divergence between the output and the `true' IMERG. Evaluated over all timesteps, this gives a distribution of KL divergences (Fig.~\ref{fig:kl_mae_combo}a) which indicates a better performance from the Tweedie model. There are two desirable properties here. Firstly, that the median of the Tweedie KL distribution is closer to zero -- indicating that more timesteps have a distribution of predicted rainfall closer to the observation than in the RMSE model. Secondly, that the right tail is considerably weaker in the Tweedie model -- indicating fewer busts.
We next compute a set of forecast metrics for each of $t+1$ through $t+4$. As the models are only trained to produce predictions for $t+1$, we use autoregression for subsequent timesteps, such that the prediction for $t+4$ comprises three prediction timesteps and one observation.
The mean absolute error over wet pixels (Fig.~\ref{fig:kl_mae_combo}b) is consistently smaller in the Tweedie model than the RMSE model. The improvement is initially modest -- from 0.452~mm at $t+1$ to 0.444~mm (a reduction of 2\%). However, by $t+4$, the reduction is considerable at 16\%.
Similarly, the hit rate for dry pixels improves by about 2\% (0.915 vs 0.896) in the Tweedie model for $t+1$, but that improvement increases to about 13\% (0.791 vs 0.701) at $t+4$.




\section{Conclusions}

We have argued that root-mean-square error (RMSE) is a poor default objective for precipitation because it imposes Gaussian residuals on a target that is not remotely Gaussian, being semi-continuous, zero-inflated, non-negative, and heavy-tailed. As a statistically supported alternative, we advocate the Tweedie deviance, a likelihood-based, differentiable loss whose variance function $V(\mu)=\mu^{p}$ admits a compound Poisson–Gamma distribution for $1<p<2$. It thus naturally accommodates the point mass at zero and the skewed positive support observed for rainfall.

Empirically, we demonstrated that precipitation across three independent datasets is far from Gaussian at all practical temporal aggregations, with Tweedie power $p$ lying firmly between the Poisson and Gamma limits, increasing with accumulation period towards the latter. Estimating $p$ via the variance–mean power law yields values that are sensitive to both timescale and location. Using the same architectures and training setup, replacing RMSE with Tweedie deviance, we:
\begin{itemize}
	\item produced comparable overall scores on downscaling but a materially better extreme recall (e.g., $\sim$0.60 vs~$\sim$0.50 at the 99th percentile in the downscaling model), and
	\item improved nowcasting in a way that compounds under autoregressive rollout: small improvements at $t{+}1$ in wet-pixel MAE and dry hit rate increased to considerably by $t{+}4$, alongside a thinner right tail in the per-timestep KL divergence distribution (fewer busts).
\end{itemize}

This matters because the training objective shapes what the network learns. A Gaussian-implied loss underpenalises heavy events, allows negative precipitation, and ignores the mass at zero. The Tweedie deviance aligns the objective with observed precipitation statistics, while remaining simple to implement. Because the deviance is continuous in $p$, it provides a single framework that smoothly bridges short-interval, arrival-dominated behaviour and longer-interval accumulation statistics.

Our results use a single global $p$ per experiment. Future work should quantify sensitivity to incorrect $p$, consider spatially or seasonally varying $p$, and compare alternative estimators (e.g., profile-likelihood) against the blockwise variance–mean fit we used here. 
We would also like to explore extending these results to probabilistic training targets, which would allow quantification of model prediction uncertainty. Finally, the observed ``weekly dip'' in the $p$–aggregation curve needs a targeted analysis to distinguish finite-sample artefacts from genuine multiscale structure.

For deterministic precipitation learning tasks, we therefore recommend replacing RMSE with Tweedie deviance. $p$ can be estimated from the training data at the target accumulation period, and non-negativity can be enforced at the output (although it may naturally arise from using the Tweedie deviance). This approach delivers equal-or-better performance overall and systematically improves behaviour where it matters most, extremes.





\appendix
\section{Model definitions}
\label{app:models}

\subsection{Common conventions}
\label{app:common}
All targets are non-negative precipitation in millimetres per model timestep (ERA5 hourly; IMERG half-hourly). To stabilise optimisation and preserve physical support we:
\begin{itemize}
	\item enforce non-negativity at the output via a smooth positive mapping (using softplus);
	\item scale inputs and targets by a single scalar \(s\), defined as the 99th percentile of the \emph{training} target pixels over space–time;
	\item use AdamW (learning rate \(2\times10^{-4}\), weight decay \(10^{-4}\)), global gradient-norm clip at 1.0, early stopping on validation loss with a patience of 4 epochs, and an exponential moving average (EMA) of weights with decay 0.999 for validation/sampling;
	\item for Tweedie deviance, set dispersion \(\phi=1\) and estimate the power \(p\) once on the \emph{training} split (for more details, see App.~\ref{app:tweedie}).
\end{itemize}

\subsection{Nowcasting model}
\label{app:nowcast}
We use a native $100\times100$ grid at 0.1$^\circ$ resolution centred on Kolkata \((22.57^\circ\mathrm{N},\,88.36^\circ\mathrm{E})\)
Our training period is 2014-06-01 to 2017-05-31; validation is 2017-06-01 to 2017-12-31; and test is 2018-06-01 to 2018-08-31.

For the LSTM, we use a context length of \(L=4\) (i.e., the last 2~h). The target is the next frame at \(t{+}1\).
Stride-1 sliding windows form examples with input tensor shape \((L,H,W,1)\) and target \((H,W,1)\) in scaled units.
Longer horizons \(t{+}k\) are obtained autoregressively by feeding the \((t{+}k{-}1)\) prediction back as the next context frame.

The architecture is as follows.
All convolutions use \(3\times3\) kernels with ``same’’ padding. The hidden width is 48 channels throughout.
Block~1 is a ConvLSTM2D layer (filters 48, return full sequence) followed by batch normalisation and SpatialDropout3D with rate 0.10.  
Block~2 is a second ConvLSTM2D layer (filters 48, return full sequence) + batch normalisation + SpatialDropout3D (also 0.10).  
A residual add combines the outputs of Blocks~1 and~2 (elementwise over the sequence).  
Finally we apply a third ConvLSTM2D (filters 48, return last frame) + batch normalisation; then a \(3\times3\) convolution with 48 filters and SiLU activation, followed by a \(1\times1\) convolution to 1 channel and a final softplus to ensure \(\mu\ge0\).
Kernels use He-normal initialisation; biases are zero. Batch-norm scales/offsets start at 1/0 as standard.

For training, we use a batch size 16 and a maximum of 50 epochs with early stopping.

\subsection{Downscaling model }
\label{app:downscale}

ERA5 hourly precipitation is used, with negatives are set to zero and non-finite values dropped.
Target (``high-resolution’’) is a native \(50\times50\) grid at \(\sim0.25^\circ\) centred on Beijing \((39.90^\circ\mathrm{N},\,116.41^\circ\mathrm{E})\).  
A 1.0\(^\circ\) conditioner is constructed by \(4\times4\) block-mean coarsening the native grid followed by bilinear interpolation back to the native grid size. A static orography (geopotential) field, standardised to zero mean and unit variance is added as a second conditioning channel.
Our training period is 1980-01-01 to 2016-12-31; validation is 2017-01-01 to 2017-12-31; and testing is over 2018-01-01 to 2018-12-31.

We train a conditional diffusion model that predicts the clean high-resolution field \(x_0\) (in scaled units) from a noisy version \(x_t\) together with a conditioner (coarse-upsampled precipitation and orography). During training a timestep \(t\in\{0,\dots,T{-}1\}\) is sampled, with \(T=1000\), linear \(\beta_t\in[10^{-4},2\times10^{-2}]\), \(\alpha_t=1-\beta_t\), and \(\bar\alpha_t=\prod_{s\le t}\alpha_s\); the forward process draws \(x_t=\sqrt{\bar\alpha_t}\,x_0+\sqrt{1-\bar\alpha_t}\,\epsilon\) with \(\epsilon\sim\mathcal N(0,I)\).

The score network is a depth-2 encoder–decoder with GroupNorm and SiLU activations and a base width of 48 channels. A scalar continuous time \(t_\mathrm{cont}\in[0,1]\) passes through a two-layer multilayer perceptron of width \(4\times48\) and the projected embedding is added (broadcast) before each normalisation.  
The structure of the encoder is a DoubleConv\((C_{\mathrm{in}}\!\to b)\) then \(2\times\) max-pool; DoubleConv\((b\!\to2b)\) then \(2\times\) pool; bottleneck DoubleConv\((2b\!\to4b)\).  
The structure of the decode is upsample\(\times2\); then concatenate skip from \(2b\); DoubleConv\((4b{+}2b\!\to2b)\); upsample\(\times2\); concatenate skip from \(b\); DoubleConv\((2b{+}b\!\to b)\); \(1\times1\) convolution to 1 channel and final softplus to obtain \(\tilde x_0\ge0\).  
Each DoubleConv is Conv\(3\times3\)\(\to\)GroupNorm(\(G\))\(\to\)SiLU\(\to\)Conv\(3\times3\)\(\to\)GroupNorm(\(G\))\(\to\)SiLU with time-embedding add-ins. \(G\) is the largest divisor of the channel count not exceeding 8 (to ensure exact divisibility). As before, kernels use He-normal initialisation; biases are zero; GroupNorm scales/offsets start at 1/0.

The channel stack used for training comprises \([x_t,\ \text{coarse-upsampled},\ \text{orography}]\) (first two in scaled units; orography standardised). The target is the clean high-resolution field \(x_0\) in scaled units.

We compare (i) root-MSE\((\tilde x_0,x_0)\) and (ii) Tweedie deviance \(d_p(x_0,\tilde x_0)\) with \(p\) from App.~\ref{app:tweedie}, both in scaled units and with a softplus applied. The batch size is 32 and we train up to 10 epochs with early stopping, maintaining an EMA (decay 0.999).

The generating model then has \(K=150\) evenly spaced integer timesteps from \(T{-}1\) to 0 and \(\eta=0\). At each chosen \(t\), we compute \(\tilde x_0\) and \(\hat\epsilon=(x_t-\sqrt{\bar\alpha_t}\,\tilde x_0)/\sqrt{1-\bar\alpha_t}\), then jump to the previous chosen step \(t'\) via \(x_{t'}=\sqrt{\bar\alpha_{t'}}\,\tilde x_0+\sqrt{1-\bar\alpha_{t'}}\,\hat\epsilon\). At \(t'=0\) return \(\tilde x_0\) and rescale by \(s\) back to mm~hr$^{-1}$.

\subsection{Estimating the Tweedie power $p$}
\label{app:tweedie}
We estimate $p$ using the variance–mean power law. For each pixel we create non-overlapping temporal blocks and fit \(\log \operatorname{Var}(Y)=c+p\,\log\mathbb{E}[Y]\) using ordinary least squares across blocks. Blocks are 30\,days long (giving for IMERG \(48\times30\) half-hourly samples and for ERA5 \(24\times30\) hourly samples). The estimate is then averaged across pixels and this value is fixed for validation and testing. The Tweedie dispersion parameter \(\phi\) only scales the deviance and is set to 1 during training.

\subsection{Evaluation metrics}
\label{app:metrics}
At each time, let \(Z\) be the set of pixels and \(\widehat y\) the prediction. Wet-pixel MAE is defined as the average \(|\widehat y_i-y_i|\) over \(i\in W=\{i\in Z: y_i\ge\tau\}\). 
Dry hit rate is defined as \(\Pr(\widehat y<\tau\,|\,y<\tau)\). 
Extreme recall is defined as \(\Pr(\widehat y\ge u_q\,|\,y\ge u_q)\), where \(u_q\) is the \(q\)th percentile of truth over the evaluation period (nowcasting \(q{=}99.9\), downscaling \(q{=}99\)).

\section*{Acknowledgments}
KMRH is supported by a NERC Independent Research Fellowship (MITRE; NE/W007924/1).

\bibliographystyle{ametsoc2014}
\bibliography{bibliography}

\end{document}